# Sensitivity to Calibrated Parameters*

Thomas H. Jørgensen†

March 11, 2021

**Abstract**

A common approach to estimation of dynamic economic models is to calibrate a sub-set of model parameters and keep them fixed when estimating the remaining parameters. Calibrated parameters likely affect conclusions based on the model but estimation time often makes a systematic investigation of the sensitivity to calibrated parameters infeasible. I propose a simple and computationally low-cost measure of the sensitivity of parameters and other objects of interest to the calibrated parameters. In the main empirical application, I revisit the analysis of life-cycle savings motives in Gourinchas and Parker (2002) and show that some estimates are sensitive to calibrations.

(JEL: C10, C52, C60)

*I wish to thank Bo Honoré, Martin Browning, John Rust, Timothy Christensen, Joan Llull, Hamish Low, Søren Leth-Petersen, Jesper Riis-Vestergaard Sørensen, Anders Munk-Nielsen, Jeppe Druedahl, Rasmus Søndergaard Pedersen, Mette Ejrnæs, Heino Bohn Nielsen, Jonathan Parker, Florian Oswald and seminar participants at CCE, GSE Structural Microeconometrics 2019 and ESEM 2019 in Manchester for helpful discussions. Center for Economic Behavior and Inequality (CEBI) is a center of excellence at the University of Copenhagen, founded in September 2017, financed by a grant from the Danish National Research Foundation, Grant DNRF134. All errors are my own.

†CEBI, Department of Economics, University of Copenhagen, Øster Farimagsgade 5, 2353 Copenhagen Denmark. E-mail: thomas.h.jorgensen@econ.ku.dk. Webpage: www.tjeconomics.com.

# 1 Introduction

Estimated dynamic economic models are now widely used across all fields of economics. The estimation of structural models is, however, notoriously time consuming and the estimation time increases drastically with the number of estimated parameters. A common approach to alleviate this computational burden is to calibrate a sub-set of the model parameters and keep them fixed while estimating the remaining parameters of interest.[1] The calibrated parameter values are often based on external sources such as previously published parameter estimates and will generally influence conclusions drawn from the estimated model. Unfortunately, a systematic investigation of the sensitivity to calibrated parameters is often infeasible. If sensitivity is investigated, the current practice is to report results from a few re-estimated versions of the model. This approach is, however, generally very time consuming and the number of alternative calibrations thus typically low, ultimately reducing research transparency. In this paper, I propose a complementary approach that can greatly improve the transparency of structural research.

I propose a low-cost measure of the sensitivity of any quantity of interest to the calibrated parameters. The sensitivity measure can often be calculated with little additional programming and without significant computational cost since it avoids re-estimation of the model parameters. The measure has a straightforward interpretation as the effect on quantities of interest from a marginal change in the calibrated parameters and can e.g. be used to construct elasticities. Like most existing types of sensitivity and robustness analyses, the proposed measure is thus local. I find the encouraging result, however, that the measure provides a quite good approximation to even larger changes in the calibrated parameters in my main empirical application.

The sensitivity measure is based on the General Method of Moments (GMM) estimation framework, which includes most commonly used estimators. I use this framework to derive a sensitivity measure that is robust to misspecification. I also provide an approximation that is particularly simple to calculate and is expected to be good for correctly specified models.

---

[1] See e.g. Gourinchas and Parker (2002); Scholz, Seshadri and Khitatrakun (2006); Cagetti and De Nardi (2006); De Nardi, French and Jones (2010); French and Jones (2011); Blundell, Dias, Meghir and Shaw (2016); Berger and Vavra (2015); Voena (2015); Chiappori, Dias and Meghir (forthcoming); Sommer (2016); Fernández and Wong (2014); Ejrnæs and Jørgensen (2020); Huo and Ríos-Rull (forthcoming) for a small sample of studies.

The sensitivity of estimated parameters or other quantities of interest can be calculated straightforwardly. The objects of primary interest are often some functions of the parameters, such as counterfactual policy reforms, optimal policy design, or welfare measures. The proposed measure can quantify the sensitivity of such model-based results to calibrations. In previous research – if theسسensitivity of such results are investigated – the most common approach is to re-calculate the objects of interest from a change in the calibrated parameters while the estimated parameters remain fixed at their originally estimated values. That approach, however, completely ignores the effect on the estimated parameters and can produce misleading results. The sensitivity measure I propose takes this effect on the estimated parameters into account without costly re-estimations of the model.

Generalizations of the estimated model can be investigated without estimating the richer and more complex model. Imagine having estimated a restricted version of a more general model class where estimating the general model is significantly more time consuming that estimating the restricted model. The sensitivity measure can be used to assess the sensitivity of results to the general model without ever having to estimate the more computationally time demanding model. The measure could also be used to construct formal Lagrange multiplier tests or provide a one-step estimate of the general model.

The sensitivity measure can greatly improve the transparency of work in many fields of economic research. I illustrate the usefulness of the approach through an application to the importance of different savings motives over the life cycle, as studied in the seminal work by Gourinchas and Parker (2002). I show how the estimation results are especially sensitive to the calibrated value of the risk-free interest rate, a parameter not considered in their original robustness analysis. I compare the proposed sensitivity measures to a brute-force re-estimation and find very encouraging results in the sense that the sensitivity measure is close to the true marginal effects of changes in the calibrated parameters. The main result in Gourinchas and Parker (2002) is that buffer-stock saving is the dominating savings motive early in working life while retirement saving or life-cycle motives are more important later in the working life. Interestingly, I find that this result is insensitive to the calibrated parameters.

Finally, I also apply the sensitivity measure to recent research on home-ownership and the option value of migration in Oswald (2019). I find that the willingness to pay for the insurance value of migration against *regional* shocks are sensitive to calibrated parameters related to *idiosyncratic* risk. This example illustrates the value of the sensitivity measure because the richness of this dynamic model in practice makes many re-estimations infeasible.

## 1.1 Existing Literature and Roadmap

The literature on sensitivity in economics is growing. Especially the sensitivity of estimators to the included moments in GMM-type estimators has received recent attention, see e.g. Kitamura, Otsu and Evdokimov (2013) and Andrews, Gentzkow and Shapiro (2017, 2020).[2] I extend the approach used by Andrews, Gentzkow and Shapiro (2017) to the current setting, investigating the sensitivity to calibrated parameters. To my knowledge, the only paper investigating the sensitivity of structural estimators to calibrated parameters is the recent paper by Iskrev (2019). That paper focuses on Bayesian approaches to estimation of macroeconomic models. I propose a measure based on a more commonly used type of estimator, derived from the minimization of a quadratic criterion, in a frequentist framework. Importantly, the approach in Iskrev (2019) requires identification of all (both calibrated and estimated) model parameters simultaneously and the availability of the covariance between the two sets of parameters. This e.g. precludes the use of external sources for calibration, which is a very common approach.[3] Another related study is by

---

[2] Analyzing local misspecification is not a new idea, however. For earlier work, see e.g. Newey (1985).

[3] A worst-case upper bound on the co-variance structure could be estimated following the approach suggested in Cocci and Plagborg-Møller (2019), however.

Chernozhukov, Escanciano, Ichimura, Newey and Robins (2018) who provide a setup for construction of *locally robust* moments that are orthogonal to the calibrated parameters. Their approach can completely eliminate sensitivity to calibrated parameters but relies on the availability of estimation information regarding the calibrated parameters. Again, this would preclude calibration based on external sources.

There are several other related studies. Bonhomme and Weidner (2018) study the sensitivity to model misspecification and provide estimators to minimize the effect on quantities of interest. Christensen and Connault (2019) study counterfactual sensitivity to assumptions about unobserved heterogeneity. Honoré, Jørgensen and de Paula (2020) propose measures related to the effect on inference from e.g. changing the weighting put on moments in estimation. Armstrong and Kolesár (2021) study the sensitivity to moments included in the estimation and propose optimal weights that can reduce the sensitivity to included moments. While all these measures are local, Harenberg, Marelli, Sudret and Winschel (2019) suggest constructing a global approximation of the object of interest to reduce the computational time required for *global* sensitivity analysis. This approach, however, still requires a significant number of re-estimations of the model.

In engineering and operations research, sensitivity and uncertainty quantification of model outputs to model inputs have received substantial attention. Some measures in this literature (see e.g. Borgonovo and Apostolakis, 2001) bears resemblance to the one, I propose, but in completely different contexts. There is also a growing focus in this literature on global measures of sensitivity, see e.g. Borgonovo and Plischke (2016) for a recent review.

The remainder of the paper is organized as follows. In the following section, I specify the estimation framework and define the sensitivity measure. In Section 3, I apply the measure to an empirical analysis of the relative importance of alternative savings motives over the life cycle. In section 4, I apply the approach to a rich model of home-ownership and migration before concluding in Section 5. Python code generating all results in this paper is available from the author's web-page.

## 2 Framework and Sensitivity

I focus on situations in which the interest lies in estimating a $K \times 1$ vector of parameters, $\theta$, given some $L \times 1$ vector of calibrated parameters, $\hat{\gamma}$. The calibrated parameters, $\hat{\gamma}$, could e.g. be estimated from different data-sources, by other researchers, or published in other papers. Interest may then be in using these estimates to subsequently analyze different model outcomes and predictions.

I assume that a GMM-type estimation approach (Hansen, 1982) is employed,

$$\hat{\theta}(\hat{\gamma}) = \arg\min_{\theta \in \Theta} g_n(\theta|\hat{\gamma})' W_n g_n(\theta|\hat{\gamma}) \tag{1}$$

where $g_n(\theta|\hat{\gamma}) = \frac{1}{n} \sum_{i=1}^n f(\theta|\hat{\gamma}, \mathbf{w}_i)$ is some $J \times 1$ vector valued function of the parameters and data, $\mathbf{w}_i$ for $i = 1, \ldots, n$, specified by the researcher. $G_n = \left.\frac{\partial g_n(\theta|\hat{\gamma})}{\partial \theta'}\right|_{\theta=\hat{\theta}}$ and $D_n = \left.\frac{\partial g_n(\hat{\theta}|\gamma)}{\partial \gamma'}\right|_{\gamma=\hat{\gamma}}$ are $J \times K$ and $J \times L$ Jacobians, respectively, and $W_n$ is a symmetric positive definite weighting matrix. I assume that *i)* $g_n(\theta|\gamma)$ is continuously differentiable in $\theta$ and $\gamma$ around $(\hat{\theta}, \hat{\gamma})$, *ii)* $G_n' W_n G_n$ has an inverse, and *iii)* $\hat{\theta}$ is in the interior of the parameter space $\Theta$, where $\Theta$ is a compact subset of $\mathbb{R}^K$, and $\hat{\theta}$ is identified in the sense that it is the unique value that solves the problem in eq. (1).

### 2.1 Current Practice to Sensitivity

In existing research, the sensitivity to calibrated parameters is often not investigated, despite the likely importance for subsequent results derived from the estimated model. If sensitivity to calibration is investigated, the current practice is to report $M$ estimation results, $\{\tilde{\theta}_m\}_1^M$, from $M$ alternative calibrations, $\{\tilde{\gamma}_m\}_1^M$, in a robustness exercise.

While this more global approach has the potential to investigate alternative relevant calibrations, it is somewhat arbitrary and rely on the researcher's priors on what might be important to investigate. A reader cannot infer the implications of her own prior from the reported robustness results if her prior is not equal to that of the researcher. I view this as a lack of transparency.

Furthermore, the current practice involves re-estimating the model $M$ times, which is potentially computationally time consuming if a dynamic economic model is solved every time the objective function is evaluated. This, in turn, often leads to a low value of $M$, reducing the transparency drastically. This motivates the low-cost complementary measure I propose below.

An example of the current practice is found in the main application below in which I revisit the seminal work of Gourinchas and Parker (2002).[4] Among other things, they investigate the sensitivity of their estimation results to changing the income shock variances and find that results are quite insensitive to these parameters. I confirm that finding using my proposed sensitivity measure. Interestingly, however, the calibrated parameter that I find to be most important (the risk-free interest rate) is not considered in their robustness exercise.

## 2.2 Sensitivity Measure

I propose the Jacobian

$$\hat{S} = \frac{\partial \hat{\theta}}{\partial \hat{\gamma}'} \tag{2}$$

as the sensitivity of the estimated parameters to the calibrated parameters. With this measure, the elasticity of the $k$th estimated parameter to the $l$th calibrated parameter can thus easily be calculated as

$$\hat{\mathcal{E}}_{(k,l)} = \hat{S}_{(k,l)} \hat{\gamma}_{(l)} / \hat{\theta}_{(k)} \tag{3}$$

assuming that $\hat{\gamma}_{(l)}, \hat{\theta}_{(k)} \neq 0$.

Proposition 1 shows how the sensitivity measure can be calculated *without* re-estimating the model. All elements can be calculated using numerical derivatives if closed form derivatives are not available.

**Proposition 1.** *The derivative of the estimated parameters w.r.t. the calibrated param-*

---

[4] A similar approach is adopted in e.g. De Nardi, French and Jones (2010); French and Jones (2011); Fernández and Wong (2014); Druedahl and Martinello (forthcoming); Jørgensen (2017); Fan, Seshadri and Taber (2019); and Druedahl and Jørgensen (2020).

*eters is the $K \times L$ matrix*

$$\frac{\partial \hat{\theta}}{\partial \hat{\gamma}'} = - \left[ (g_n(\hat{\theta}(\hat{\gamma})|\hat{\gamma})' W_n \otimes I_K) C_{\theta,n} + G_n' W_n G_n \right]^{-1}$$
$$\times \left[ (g_n(\hat{\theta}(\hat{\gamma})|\hat{\gamma})' W_n \otimes I_K) C_{\gamma,n} + G_n' W_n D_n \right] \quad (4)$$

*if the inverse exists, where $C_{\theta,n} = \left. \frac{\partial vec(G_n')}{\partial \theta'} \right|_{\theta=\hat{\theta}}$ is a $JK \times K$ matrix of stacked second order derivatives and $C_{\gamma,n} = \left. \frac{\partial vec(G_n')}{\partial \gamma'} \right|_{\gamma=\hat{\gamma}}$ is a $JK \times L$ matrix with stacked cross-derivatives.*

*Proof.* See the Supplemental Material Section A. □

Corollary 1 shows an approximation to $\frac{\partial \hat{\theta}}{\partial \hat{\gamma}'}$ that is particularly fast to calculate because almost all elements are already constructed when calculating the asymptotic covariance matrix of $\hat{\theta}$.[5] Calculating the only missing element, $D_n$, involves only $2L$ evaluations of the objective function if central finite differences are used to calculate numerical derivatives. The condition for a good approximation is satisfied in just-identified cases since $g_n(\hat{\theta}(\hat{\gamma})|\hat{\gamma}) \approx 0$ in that case. In correctly specified over-identified models ($J > L$) we might also have $g_n(\hat{\theta}(\hat{\gamma})|\hat{\gamma})' W_n \approx 0$. Generally, however, the approximation is expected to be good if potential non-zero moments are approximately linear in parameters around $(\hat{\theta}, \hat{\gamma})$.

**Corollary 1.** *If $(g_n(\hat{\theta}(\hat{\gamma})|\hat{\gamma})' W_n \otimes I_K) C_{\theta,n} \approx 0_{K \times K}$ and $(g_n(\hat{\theta}(\hat{\gamma})|\hat{\gamma})' W_n \otimes I_K) C_{\gamma,n} \approx 0_{K \times L}$ the derivative can be approximated as*

$$\hat{S} \approx \Lambda_n D_n \quad (5)$$

*where $\Lambda_n = -(G_n' W_n G_n)^{-1} G_n' W_n$.*

Under standard regularity conditions – which imply that $\hat{\theta}$ is a consistent estimator of the true population value – the approximation in Corollary 1 is a consistent estimator of $\frac{\partial \theta}{\partial \gamma'}$ in the population (see Supplemental Material B). In the main empirical application below, I compare the approximation to the robust formula and the brute-force derivative using (costly) re-estimation of the model and find that the approximate sensitivity measure in eq. (5) is a quite good approximation of $\frac{\partial \hat{\theta}}{\partial \hat{\gamma}'}$.

---

[5] In fact, if asymptotic standard errors are corrected for the two-step estimation approach, as in Gourinchas and Parker (2002), all elements of the sensitivity measure is already calculated.

**Example** (Linear Regression). Consider a simple linear regression model with two mean-zero explanatory variables, $X_1$ and $X_2$, and measurement error, $\varepsilon$,

$$Y_i = \beta_1 X_{1,i} + \beta_2 X_{2,i} + \varepsilon_i$$

where $\mathbb{E}[\varepsilon|X_1, X_2] = 0$ is the identifying assumption. Imagine fixing the second parameter to $\hat{\beta}_2$ and only estimating $\beta_1$ with a single moment in $g_n(\beta_1|\hat{\beta}_2) = \frac{1}{n}\sum_{i=1}^n (Y_i - \beta_1 X_{1,i} - \hat{\beta}_2 X_{2,i})X_{1,i}$. This ordinary least squares (OLS) estimator can be found in closed form as

$$\hat{\beta}_1(\hat{\beta}_2) = \frac{\sum_{i=1}^n X_{1,i}(Y_i - \hat{\beta}_2 X_{2,i})}{\sum_{i=1}^n X_{1,i}^2}. \tag{6}$$

In this setting, $G_n = -\sum_{i=1}^n X_{1,i}^2$ and $D_n = -\sum_{i=1}^n X_{1,i}X_{2,i}$ and the sensitivity measure is

$$\hat{S} = -\frac{\sum_{i=1}^n X_{1,i}X_{2,i}}{\sum_{i=1}^n X_{1,i}^2}$$

showing the intuitive result that if $X_1$ and $X_2$ are positively (negatively) correlated, increasing $\hat{\beta}_2$ would lead to a reduced (increased) $\hat{\beta}_1$. If they are uncorrelated, the estimator of $\beta_1$ is insensitive to $\hat{\beta}_2$. In this just-identified example, it is also easy to verify that $\hat{S} = \frac{\partial \hat{\beta}_1(\hat{\beta}_2)}{\partial \hat{\beta}_2}$ and that $-\hat{S}\hat{\beta}_2$ is equal to the omitted variable bias from omitting $X_2$ in the OLS regression.

The sensitivity measure is related to that proposed by Andrews, Gentzkow and Shapiro (2017). In particular, they propose to report $\hat{\Lambda}$ as a local measure of the sensitivity of $\theta$ to the included estimation moments in $g_n(\bullet)$. They do not consider the topic of the current paper and thus do not discuss sensitivity to calibrated parameters. The measure that I propose addresses this by weighting $\hat{\Lambda}$ by the effect of the calibrated parameters on each included moment through $D_n$ in (5). In the Supplemental Material, I discuss other related sensitivity measures.

## 2.3 Extensions

**Sensitivity of other Quantities of Interest.** Denote a $F \times 1$ vector of quantities of interest as $h(\theta, \gamma) = \frac{1}{n} \sum_{i=1}^{n} h_i(\theta, \gamma | \mathbf{w}_i)$. Examples include counterfactual policy simulations and welfare measures. The proposed measure can easily be used to construct a sensitivity measure of such quantities of interest using the chain rule. The sensitivity of a $F \times 1$ vector of statistics, $h(\theta, \gamma)$, to the calibrated parameters is given by the $F \times L$ matrix

$$\hat{H} = A_n + B_n \hat{S} \tag{7}$$

where $A_n = \left.\frac{\partial h(\hat{\theta}, \gamma)}{\partial \gamma'}\right|_{\gamma=\hat{\gamma}}$ is a $F \times L$ Jacobian matrix of $h(\theta, \gamma)$ w.r.t. $\gamma$ and $B_n = \left.\frac{\partial h(\theta, \hat{\gamma})}{\partial \theta'}\right|_{\theta=\hat{\theta}}$ is a $F \times K$ Jacobian matrix of $h(\theta, \gamma)$ w.r.t. $\theta$. Again, $\hat{A}$ and $\hat{B}$ are often relatively fast to calculate numerically.

**Sensitivity to Arbitrary Changes in $\gamma$.** The sensitivity to an arbitrary change in $\gamma$, say $\Delta\gamma = \tilde{\gamma} - \hat{\gamma}$ could be constructed as a sum of the individual derivatives (see e.g. Borgonovo and Apostolakis, 2001). Denote the $K \times 1$ vector of sensitivity measures of $\theta$ to the $j$th element in $\gamma$ as $\hat{S}_{(j)}$. The sensitivity of $\theta$ to a change $\Delta\gamma$ could then be calculated as

$$\hat{S}_{\Delta\gamma} = \sum_{j=1}^{J} \hat{S}_{(j)} \Delta\gamma_{(j)}. \tag{8}$$

Importantly, these measures can be calculated by readers applying their own priors if only the sensitivity measure $\hat{S}$ is reported.

**Sensitivity to Generalizations.** In the same spirit as Lagrange multiplier tests, the sensitivity measure can be used to investigate general versions of the estimated model. The idea is again to avoid estimating the general model which can be significantly more time consuming to estimate than the restricted model.

To formalize this idea, partition $\gamma$ into $\gamma = (\gamma_1, \gamma_2)$. Nested in the general model is the estimated model with $\gamma_2 = 0$ and associated moment function $g_n(\theta|\gamma_1, 0)$. The sensitivity to a general version of the model is thus given by eq. (5) with $D_n = \left.\frac{\partial g_n(\hat{\theta}|\hat{\gamma}_1, \gamma_2)}{\partial \gamma_2}\right|_{\gamma_2=0}$. The general model must be used when constructing the sensitivity measure but this does

not involve re-estimation. The sensitivity measure could even be used to form a one-step estimate of the generalized model.

## 3   Application: Life-Cycle Savings Motives

In a seminal paper, Gourinchas and Parker (2002) estimate a dynamic structural model of life-cycle consumption and saving using data for the US. They use the estimated model to study the importance of life-cycle (retirement) and buffer (risk) related motives for saving over the life-cycle. Here, I illustrate the usefulness of the sensitivity measure through my implementation of that analysis.

The recursive form of the model is

$$V_t(M_t, P_t) = \max_{C_t \in (0, M_t]} v_t \frac{C_t^{1-\rho}}{1-\rho} + \beta \mathbb{E}_t[V_{t+1}(M_{t+1}, P_{t+1})]$$

s.t.

$$M_{t+1} = (1+r)(M_t - C_t) + Y_{t+1}$$

$$Y_{t+1} = P_{t+1} U_{t+1}$$

$$P_{t+1} = G_{t+1} P_t N_{t+1}$$

$$\log N_{t+1} \sim \mathcal{N}(0, \sigma_n^2)$$

$$U_{t+1} = \begin{cases} \tilde{U}_{t+1} & \text{with probability } 1-p \\ 0 & \text{with probability } p \end{cases}$$

$$\log \tilde{U}_{t+1} \sim \mathcal{N}(0, \sigma_u^2)$$

for $t \leq T$ where $\beta$ is the discount factor and $\rho$ is the coefficient of constant relative risk aversion (CRRA). The initial wealth, $W_{26} = (1+r)(M_{25} - C_{25})$, is drawn from a log-normal distribution such that $\log W_{26} \sim \mathcal{N}(\omega_{26}, \sigma_{\omega_{26}}^2)$ and initial permanent income is $P_{26}$ for all households. Consumers face log-normal permanent and transitory income shocks, denoted $N_t$ and $U_t$, respectively. Furthermore, consumers experience a transitory zero-income shock with probability $p$. The income growth factor, $G_t$, and taste shifter associated with family composition, $v_t$, evolves deterministically and are perfectly foreseeable by consumers. At retirement, a simple linear consumption function is assumed to apply, $c_{T+1} = \gamma_0 + \gamma_1 m_{T+1}$ where $c_{T+1} = C_{T+1}/P_{T+1}$ and $m_{T+1} = M_{T+1}/P_{T+1}$ are normalized consumption and marked resources, respectively. Further details of the economic model as well as the numerical solution approach is given in the Supplemental Material and in the original paper.

The authors estimate $\theta = (\beta, \rho, \gamma_0, \gamma_1)$ and keep all other parameters (which I denote $\gamma$) fixed at calibrated values using simulated minimum distance (SMD). The authors include log-average consumption from age 26 through 65 as moments and use as preferred weight a diagonal matrix with the inverse of the variance of the empirical moments on the diagonal.[6]

The calibrated parameters are reported in Table S1 and Figure S1 in the Supplemental Material where $\tilde{\omega}_{26} = \exp(\omega_{26})$. I use similar calibrations as Gourinchas and Parker (2002) but I re-estimate $\beta$ and $\rho$ using my implementation while fixing $\gamma_0 = 0.0015$ and $\gamma_1 = 0.071$ to the estimated values in Gourinchas and Parker (2002). I use the re-estimated model throughout with $\hat{\beta} = 0.944$ and $\hat{\rho} = 1.860$. The model fit is illustrated in Figure S2.

## 3.1 Sensitivity of Parameter Estimates

Table 1 reports the sensitivity measure (in elasticities) in columns 2–5 together with brute-force elasticities based on re-estimation of the model in columns 6–7. I use central finite differences when constructing sensitivity measures while the brute-force calculated effect on the $k$th element of $\theta$ of a $\epsilon$ percent increase in the $l$th element in $\gamma$ is calculated as $(\tilde{\theta}^l_{(k)} - \hat{\theta}_{(k)})/\hat{\theta}_{(k)} \cdot 100$ where $\tilde{\theta}^l_{(k)}$ is the $k$the element in $\tilde{\theta}^l = \arg\min_{\theta \in \Theta} g_n(\theta|\tilde{\gamma}^l)'W_n g_n(\theta|\tilde{\gamma}^l)$ with $\tilde{\gamma}^l = \hat{\gamma}(1_L + \iota_l \cdot \epsilon/100)$. The approximate sensitivity measures in Table 1 was calculated in roughly 0.7 minutes while the robust measure was calculated in roughly 3

---

[6] For each value of $\theta$, I solve the model using the endogenous grid method (EGM), proposed by Carroll (2006), rather than time iteration used in Gourinchas and Parker (2002). The EGM is faster and more accurate than time iteration (see e.g. Jørgensen, 2013). The Supplemental Material contains a detailed description of the implementation.

additional minutes. In contrast, the brute force measures was calculated in roughly 17 minutes.[7]

All sensitivity measures are close to the brute-force calculations without being based on re-estimations of the model. This suggests that the approximate sensitivity measure is a quite good approximation in this application. Recall that the sensitivity measures are local (derivatives) while the brute-force calculations are based on a one percent increase in $\hat{\gamma}$ partly explaining the observed differences. In the supplemental material (Figure S3), I compare the approximate derivative with the robust and brute-force derivatives and find that all three are quite similar.

I find that the discount rate is relatively insensitive to the calibrated parameters while the CRRA coefficient, $\rho$, is sensitive to all calibrated parameters. As also found in Gourinchas and Parker (2002), I find that lowering the transitory and permanent income shock variances have minuscule effects on the value of the estimated discount factor and only slightly affect the estimated risk aversion coefficient.

Table 1: Sensitivity of Parameters. Elasticities.

|  | Sensitivity measure | | | | Re-estimation | |
|---|---|---|---|---|---|---|
|  | Approximation | | Robust | | (brute force) | |
|  | $\hat{\beta}$ | $\hat{\rho}$ | $\hat{\beta}$ | $\hat{\rho}$ | $\hat{\beta}$ | $\hat{\rho}$ |
| $\sigma_n$ | -0.001 | -0.023 | -0.002 | 0.041 | -0.003 | 0.055 |
| $\sigma_u$ | 0.001 | -0.069 | 0.000 | -0.025 | 0.001 | -0.063 |
| $p$ | 0.009 | -0.359 | 0.011 | -0.436 | 0.009 | -0.408 |
| $r$ | -0.001 | -1.365 | -0.016 | -0.687 | -0.010 | -0.945 |
| $\tilde{\omega}_{26}$ | -0.010 | 0.435 | -0.009 | 0.369 | -0.010 | 0.413 |
| $\sigma_{\omega_{26}}$ | -0.016 | 0.670 | -0.010 | 0.503 | -0.013 | 0.599 |

*Notes*: The table reports the sensitivity of the estimated parameters in $\theta$ to the calibrated parameters in $\gamma$. The left panel reports the proposed sensitivity measure as elasticities. The right panel shows the same statistics calculated "brute-force" as the percentage change relative to the baseline with re-estimated $\theta$ parameters.

The CRRA coefficient is particularly sensitive to the probability of a zero-income shock, $p$, the initial wealth distribution, $\tilde{\omega}_{26} = \exp(\omega_{26})$ and $\sigma_{\omega_{26}}$, and the risk-free interest rate, $r$. Like the CRRA coefficient, the zero-income shock probability, $p$, affects the curvature of the consumption function for lower levels of resources (see, e.g. the discussion in Carroll, 1992, 1997). Increasing either $\rho$ or $p$ would tend to lower consumption for low levels of resources. In turn, if $p$ is increased, $\rho$ would have to decrease to match the observed consumption profile. On the other hand, if the mean initial level of wealth is increased either through an increase in $\tilde{\omega}_{26}$ or $\sigma_{\omega_{26}}$, the CRRA coefficient, $\hat{\rho}$, would increase to maintain the fit of the observed consumption profile. Such a positive relationship is also found in Gourinchas and Parker (2002).

---

[7] All timings was done on a Lenovo laptop with 4 Intel(R) i7-8665U CPUs @ 1.90GHz and 16GB RAM and should be seen as illustrative. Timings can differ significantly across models and implementations.

The parameter to which the estimates are most sensitive (in percentage terms) is the risk-free interest rate, $r$, with an elasticity of around $-1.4$. This parameter is not varied in the original study. The sign is negative because increasing the risk-free interest rate increases the value of holding wealth through a dominating substitution effect, decreasing consumption.[8] The same is true with the CRRA coefficient. In turn, increasing the interest rate will lead to a reduction in $\hat{\rho}$ in order to match the consumption age profile in the data.

Table 2 investigates the sensitivity to larger increases in the risk-free interest rate from one to five percent. The table shows the sensitivity measure elasticities from (3) in the top panel through linear extrapolation together with the actual (brute-force) percentage change in the estimated parameter values in the bottom panel. The latter brute-force approach requires re-estimation of the model for each new value of $r$ but measures the "true" marginal effects of the larger interest changes considered. Since the sensitivity measure is local and calculated at the baseline $r$, one would expect it to be a better approximation for small changes. This is confirmed by a very small difference between the sensitivity measure and the brute-force changes to a one-percent increase in $r$. Overall, however, the sensitivity measure performs very well and deviations are relatively small for even larger changes in $r$.

---

[8] See e.g. the discussion in Carroll, Slacalek and Sommer (2019).

Table 2: Sensitivity of Estimates to Large Changes in $r$. Percentages.

| | \multicolumn{5}{c}{Change in interest rate, $r$} | | | | |
|---|---|---|---|---|---|
| | 1 pct. | 2 pct. | 3 pct. | 4 pct. | 5 pct. |
| | \multicolumn{5}{c}{*Sensitivity measure (Approximate)*} | | | | |
| $\hat{\beta}$ | -0.001 | -0.002 | -0.003 | -0.004 | -0.005 |
| $\hat{\rho}$ | -1.365 | -2.731 | -4.096 | -5.462 | -6.827 |
| | \multicolumn{5}{c}{*Sensitivity measure (Robust)*} | | | | |
| $\hat{\beta}$ | -0.016 | -0.032 | -0.049 | -0.065 | -0.081 |
| $\hat{\rho}$ | -0.687 | -1.374 | -2.062 | -2.749 | -3.436 |
| | \multicolumn{5}{c}{*Re-estimated $\theta$ (brute force)*} | | | | |
| $\hat{\beta}$ | -0.010 | -0.023 | -0.033 | -0.051 | -0.063 |
| $\hat{\rho}$ | -0.945 | -1.774 | -2.696 | -3.338 | -4.225 |

*Notes*: The table reports the sensitivity of $\theta$ to the risk-free interest rate, $r$. In the top panel is the proposed sensitivity measure reported as change in percent. In the bottom panel, the brute-force percentage increase relative to the baseline $r$ in estimated parameters are reported. All other parameters are fixed at their calibrated/estimated values.

## 3.2 Sensitivity of Savings Motives.

A key result in Gourinchas and Parker (2002) is the decomposition of the saving motives over the life cycle. In particular, the estimated model suggests that before age 40, households save predominantly to buffer against income shocks, referred to as buffer savings. In the remaining working life until age 65, the primary savings motive is to sustain a desired consumption level in retirement, referred to as life-cycle savings. To investigate the sensitivity of this result, I construct a measure of the difference in the two savings-motives, following the approach in Gourinchas and Parker (2002).

Let $s_{30}^{LC}$ denote the average saving due to life-cycle motives at age 30 and let $s_{30}^{B}$ be the savings due to buffer motives at age 30. I then calculate the difference $h_{30} = s_{30}^{B} - s_{30}^{LC}$ and similarly at age 60, $h_{60}$. Table 3 shows the elasticities of these statistics with respect to the calibrated parameters. The details on how these measures are constructed is included in the Supplemental Material.

Table 3: Sensitivity of Saving Motives. Elasticities.

|  | Sensitivity measure | | | | Re-estimation | |
|---|---|---|---|---|---|---|
|  | Approximation | | Robust | | (brute force) | |
|  | $H_{30}^e$ | $H_{60}^e$ | $H_{30}^e$ | $H_{60}^e$ | $H_{30}^e$ | $H_{60}^e$ |
| $\sigma_n$ | 0.046 | -0.008 | 0.045 | -0.008 | 0.046 | -0.008 |
| $\sigma_u$ | 0.011 | 0.000 | 0.014 | 0.002 | 0.012 | 0.001 |
| $p$ | -0.000 | 0.009 | -0.009 | 0.005 | 0.013 | 0.012 |
| $r$ | 0.011 | -0.120 | -0.001 | -0.119 | -0.012 | -0.127 |
| $\tilde{\omega}_{26}$ | 0.006 | -0.011 | 0.000 | -0.014 | 0.003 | -0.012 |
| $\sigma_{\omega_{26}}$ | -0.006 | -0.046 | -0.048 | -0.065 | -0.039 | -0.061 |

*Notes*: The table reports the sensitivity of the difference between the level of buffer and life-cycle savings at age 30 and 60. The left panel reports the proposed sensitivity measure as elasticities. The right panel shows the same statistics calculated as the percentage change relative to the baseline with re-estimated $\theta$ parameters.

The savings motives decomposition is rather insensitive to the calibrated parameters. While the estimated parameters are sensitive to e.g. the interest rate, the effect of the changed interest rate on the savings motives are counter-balanced by the adjustment in $\theta$ from the change in $\gamma$. In the current application, this happens to such a degree that the savings motive decomposition is hardly affected by the calibrated parameters. The reason for this is likely that the age profile of consumption (mirror of savings) is included in the estimation moments. In turn, roughly speaking, the estimator basically adjusts the estimated parameters to changes in $\gamma$ as to leave the savings profile unaffected.

The low-cost sensitivity measure is almost identical to the brute-force elasticities, calculated from re-estimating the model. This is very encouraging because it suggests that, at least in the current application, the proposed sensitivity measure has the potential to capture the complex effects on $h(\bullet)$ from changing $\gamma$ through the direct effect ($A$) and the indirect effect ($B \cdot S$) without having to re-estimate the model.

Table 4 shows the sensitivity measure of the savings-motives from larger changes in the risk-free interest rate, $r$. I include the brute-force re-estimation results with an alternative measure in the bottom panel based on the change in the savings motives from changing $r$ while keeping $\hat{\theta}$ fixed at their baseline estimated values. This latter statistic is sometimes reported as a low-cost analysis of the sensitivity to calibrated parameters. I denote this measure as $B^e$ since it is closely related to $B$ in (7).

Again, the results are extremely encouraging. The sensitivity measure is very close to the "true" brute-force percentage changes, even for larger interest rate increases. On the other hand, the bottom panel shows that results from changing $r$ while keeping $\hat{\theta}$ fixed leads to significant *overestimation* of the effect of a change in the interest rate. This is because, unlike the sensitivity measure, this latter measure does not take into account that $\hat{\theta}$ will adjust to such a change in the calibration and thus also affect the calculated statistics.

Table 4: Sensitivity of Saving Motives to Large Changes in $r$. Percentages.

|  | \multicolumn{5}{c}{Change in interest rate, $r$} |  |  |  |
|---|---|---|---|---|---|
|  | 1 pct. | 2 pct. | 3 pct. | 4 pct. | 5 pct. |
| \multicolumn{6}{c}{*Sensitivity measure (Approximate)*} |
| $H^e_{30}$ | 0.011 | 0.022 | 0.034 | 0.045 | 0.056 |
| $H^e_{60}$ | -0.120 | -0.239 | -0.359 | -0.479 | -0.599 |
| \multicolumn{6}{c}{*Sensitivity measure (Robust)*} |
| $H^e_{30}$ | -0.001 | -0.001 | -0.002 | -0.002 | -0.003 |
| $H^e_{60}$ | -0.119 | -0.239 | -0.358 | -0.478 | -0.597 |
| \multicolumn{6}{c}{*Re-estimated $\theta$ (brute force)*} |
| $H^e_{30}$ | -0.012 | -0.020 | -0.029 | -0.016 | -0.025 |
| $H^e_{60}$ | -0.127 | -0.254 | -0.384 | -0.505 | -0.639 |
| \multicolumn{6}{c}{*Fixed $\theta$*} |
| $B^e_{30}$ | -1.005 | -2.004 | -2.998 | -3.987 | -4.971 |
| $B^e_{60}$ | -0.523 | -1.037 | -1.542 | -2.040 | -2.528 |

*Notes*: The table reports the sensitivity of the difference between the level of buffer and life-cycle savings at age 30 and 60. The top panel reports the proposed sensitivity measure as percent changes. The middle panel shows the same statistics calculated as the percentage change relative to the baseline with re-estimated $\theta$ parameters for the various values of $r$. The bottom panel illustrates the percentage change in the statistics from the change in $r$ while keeping $\theta$ fixed at their baseline estimated values.

# 4   Application: Home-ownership and the Option Value of Regional Migration

The previous application facilitated a direct comparison of the sensitivity measure with a more brute-force approach. In this second application, I illustrate the usefulness of the measure by applying it to a rich model that requires significant computational time to solve and estimate.[9] In turn, the brute-force approach is for all practical purposes infeasible.

Motivated partly by the empirical fact that homeowners migrate less than renters, Oswald (2019) estimates a rich dynamic programming model of home-ownership and migration. He then uses the estimated model to show that although the frequency of migration among homeowners is relatively low, they still value the migration option because this option acts as an insurance against adverse regional shocks.

In the model, individuals choose in which region to live, $d_j \in D$, where $j$ denotes age. Simultaneously, they choose whether to own a house, $h_j \in \{0, 1\}$ and how much to consume, $c_j$, and thus how much wealth to carry over to the following period, $a_{j+1}$. I refer the reader to the original paper for a detailed description of the model and give only a brief outline of the model here. Individuals make their optimal choices taking into account 10 state variables in $x_j = (a_j, z_j, s_j, \mathbf{F}_j, h_{j-1}, d_{j-1}, \tau, j)$ denoting, respectively, assets, an individual income shock, household size, an aggregate 2-dimensional price vector, housing status coming into the current period, the current region index, time-invariant moving cost type, and age.[10]

Oswald (2019) fixes $L = 8$ calibrated parameters in $\gamma = (\tilde{\gamma}, \beta, \rho, \sigma, \phi, \chi, r, r^m)$ and estimates $K = 19$ parameters in $\theta$ by SMD using set of $J = 38$ moments and a diagonal weighting matrix. Both sets of parameters are reproduced in Table S2 in the Supplemental Material. The estimated parameters are especially sensitive to the values of the first four parameters (reported in Table S3 in the Supplemental Material): The value of the risk aversion parameter, $\tilde{\gamma}$, the discount factor, $\beta$, the persistence of income shocks, $\rho$, and the standard error of idiosyncratic income shocks, $\sigma$. Interestingly, the parameter adjusting the continuation value of a house at the terminal period, $\omega$, and the share of high-types, $\pi_\tau$, seem relatively insensitive to most calibrated parameters.

---

[9] I am grateful to Florian Oswald for supplying $A_n$, $B_n$, $D_n$, $G_n$, and $W_n$ for his application.

[10] Oswald (2019) allows for cohort effects and indices in the original paper thus has a time-dimension denoted by $t$. For ease of exposition, I abstract from that here.

Oswald (2019) uses the estimated model to calculate the option value of migration. This exercise yields an estimated option value of migration of around $\hat{\Delta} = (\hat{\delta} - 1) \cdot 100 = 19.2\%$ (Oswald, 2019, Table 12). Table 5 illustrates the sensitivity of this measure to calibrated parameters. All numbers are elasticities based on the approximation in eq. (5). Interestingly, the option value of migration is clearly most sensitive to calibrated parameters related to risk: A one-percent increase in the risk aversion coefficient would increase the option value of migration with around one percent and a one-percent increase in the persistence of income shocks would decrease the option value with around half a percent.

Table 5: Sensitivity of the Option Value of Migration. Elasticities.

|  | $\tilde{\gamma}$ | $\beta$ | $\rho$ | $\sigma$ | $\phi$ | $\chi$ | $r$ | $r^m$ |
|---|---|---|---|---|---|---|---|---|
| $\hat{\Delta}^e$ | 1.349 | -0.127 | -0.524 | -0.026 | 0.005 | -0.053 | -0.002 | 0.002 |

*Notes*: The table reports the sensitivity of the estimated option value of migration, $\delta$, in Oswald (2019). Elasticities are reported.

The main component in the option value of migration is insurance against adverse *regional* shocks. If a consumer in the model is more risk averse, insurance against such risk is more valuable. On the other hand, if the consumer face greater *idiosyncratic* income risk due to more persistent income shocks (larger $\rho$), such an insurance mechanism is relatively *less* valuable. The reason is that increased idiosyncratic income risk will lead to increased savings in order to buffer against this risk. With more buffer-stock savings, the consumer will also have more self-insurance against adverse *regional* shocks. In turn, the option value of migration would be lower.

# 5   Concluding Discussion

A standard approach to estimation of dynamic economic models is to calibrate a sub-set of the model parameters and keep them fixed while estimating the remaining parameters of interest. If the importance of such calibrations is investigated, it is now standard to re-estimate the model using a few permutations of the calibrated parameters. In this paper, I propose an alternative approach to this relative time-consuming approach that is applicable to most popular estimators. The sensitivity measure is simple and fast to implement, yet offers an easy interpretation of the sensitivity of any quantity of interest to the calibrated parameters. In turn, the proposed sensitivity measure can greatly improve the transparency of structural research.

Applying the proposed measure to the seminal work by Gourinchas and Parker (2002) of savings motives over the life cycle, I illustrate the usefulness of the measure. The authors report re-estimated parameters varying a set of fixed parameters but do not consider e.g. the effect of the fixed risk-free interest rate. I find that especially the point estimate of the constant relative risk aversion is sensitive to several calibrated parameters – especially the risk-free interest rate. While the sensitivity measure is a local measure, the main application shows very encouraging results in the sense that the low-cost sensitivity measure is very close to the "true" brute-force effects from re-estimating the model. The same is true for an approximate version of the measure that is particularly simple to calculate.

Newey, W. K. (1985): "Generalized method of moments specification testing," *Journal of Econometrics*, 29(3), 229–256.

Oswald, F. (2019): "The Effect of Homeownership on the Option Value of Regional Migration," *Quantitative Economics*, 10(4), 1453–1493.

Scholz, J. K., A. Seshadri and S. Khitatrakun (2006): "Are Americans Saving 'Optimally' for Retirement?," *Journal of Political Economy*, 114(4), 607–643.

Sommer, K. (2016): "Fertility Choice in a Life Cycle Model with Idiosyncratic Uninsurable Earnings Risk," *Journal of Monetary Economics*, 83, 27–38.

Voena, A. (2015): "Yours, Mine, and Ours: Do Divorce Laws Affect the Intertemporal Behavior of Married Couples?," *American Economic Review*, 105(8), 2295–2332.

# Supplemental Material: Sensitivity to Calibrated Parameters

Thomas H. Jørgensen

March 11, 2021

## A  Proof of Proposition 1

Consider the problem
$$\hat{\theta}(\hat{\gamma}) = \arg\min_{\theta \in \Theta} g_n(\theta|\hat{\gamma})' W_n g_n(\theta|\hat{\gamma})$$

where $g_n(\theta|\hat{\gamma}) = \frac{1}{n}\sum_{i=1}^n f(\theta|\hat{\gamma}, \mathbf{w}_i)$ is some $J \times 1$ vector valued function of the parameters and data, $\mathbf{w}_i$ for $i = 1, \ldots, n$, specified by the researcher. $G_n = \left.\frac{\partial g_n(\theta|\hat{\gamma})}{\partial \theta'}\right|_{\theta=\hat{\theta}}$ and $D_n = \left.\frac{\partial g_n(\hat{\theta}|\gamma)}{\partial \gamma'}\right|_{\gamma=\hat{\gamma}}$ are $J \times K$ and $J \times L$ Jacobians, respectively, and $W_n$ is a symmetric positive definite weighting matrix.

The sensitivity measure that I propose to report is the change in the estimated parameters from a marginal change in the calibrated parameters, $\frac{\partial \hat{\theta}(\hat{\gamma})}{\partial \hat{\gamma}'}$. To derive this quantity in the current setup, I apply the Implicit Value Theorem to the first order condition (FOC) associated with a solution to the problem above. The FOC is

$$G_n' W_n g_n(\hat{\theta}(\hat{\gamma})|\hat{\gamma}) = 0_{K \times 1}$$

and total differentiation of the left hand side of the FOC gives

$$\frac{d}{d\hat{\gamma}'} G_n' W_n g_n(\hat{\theta}(\hat{\gamma})|\hat{\gamma}) = (g_n(\hat{\theta}(\hat{\gamma})|\hat{\gamma})' W_n \otimes I_K)[C_{\theta,n} \hat{S} + C_{\gamma,n}] + G_n' W_n [G_n \hat{S} + D_n]$$

where $\hat{S} = \frac{\partial \hat{\theta}(\hat{\gamma})}{\partial \hat{\gamma}'}$ is the object of interest, $C_{\theta,n} = \left.\frac{\partial vec(G_n')}{\partial \theta'}\right|_{\theta=\hat{\theta}}$ is a $JK \times K$ matrix of stacked second order derivatives and $C_{\gamma,n} = \left.\frac{\partial vec(G_n')}{\partial \gamma'}\right|_{\gamma=\hat{\gamma}}$ is a $JK \times L$ matrix with stacked cross-derivatives.[1] Isolating gives

$$\hat{S} = -[(g_n(\hat{\theta}(\hat{\gamma})|\hat{\gamma})'W_n \otimes I_K)C_{\theta,n} + G_n'W_nG_n]^{-1}[(g_n(\hat{\theta}(\hat{\gamma})|\hat{\gamma})'W_n \otimes I_K)C_{\gamma,n} + G_n'W_nD_n]. \tag{1}$$

# B  Consistency of the Approximate Sensitivity Estimator in Corollary 1

Under similar assumptions as those imposed in e.g. Newey and McFadden (1994) that ensure that $\hat{\theta}$ is a consistent estimator of the population value, $\theta_0$, the approximation converges in probability to the population counterpart,

$$\text{plim}_{n \to \infty} \Lambda_n D_n = S.$$

To show this, I assume that *i)* $W_n$ converges in probability to a positive semi-definite matrix $W$, *ii)* $g_n(\theta|\gamma)$, $G_n(\theta|\gamma)$ and $D_n(\theta|\gamma)$ converge uniformly in probability to their continuous population counterparts, $g(\theta|\gamma) = \mathbb{E}[f(\theta|\gamma, \mathbf{w}_i)]$, $G(\theta|\gamma) = \mathbb{E}\left[\frac{\partial f(\theta|\gamma, \mathbf{w}_i)}{\partial \theta'}\right]$ and $D(\theta|\gamma) = \mathbb{E}\left[\frac{\partial f(\theta|\gamma, \mathbf{w}_i)}{\partial \gamma'}\right]$, respectively, iii) $g(\theta|\gamma)$ is continuously differentiable in $(\theta, \gamma)$ *iv)* $\hat{\gamma}$ converges in probability to $\gamma_0$, *v)* $G'WG = G(\theta_0|\gamma_0)'WG(\theta_0|\gamma_0)$ has an inverse, and *v)* $\theta_0$ is in the interior of $\Theta$, where $\Theta$ is a compact subset of $\mathbb{R}^K$, and $\theta_0$ is identified in the sense that it is the unique value that solves $g(\theta_0|\gamma_0) = 0$. The latter assumption implicitly assumes that the model is correctly specified.

Consider the population problem

$$\theta(\gamma) = \arg\min_{\theta \in \Theta} g(\theta|\gamma)'Wg(\theta|\gamma)$$

where $\gamma = \text{plim}_{n \to \infty} \hat{\gamma}$ with associated first order condition (FOC), assuming interchangeability between integration and differentiation,

$$G(\theta(\gamma)|\gamma)'Wg(\theta(\gamma)|\gamma) = 0.$$

Using that $\text{plim}_{n \to \infty} \hat{\gamma} = \gamma_0$ and $\text{plim}_{n \to \infty} \hat{\theta} = \theta(\gamma_0) = \theta_0$ and thus $g(\theta(\gamma)|\gamma) = g(\theta_0|\gamma_0) = 0$, total differentiation of the left hand side of the FOC gives

$$\frac{d}{d\gamma'}G(\theta(\gamma)|\gamma)'Wg(\theta(\gamma)|\gamma) = G'W[GS + D] \tag{2}$$

where $S = \frac{\partial \theta(\gamma)}{\partial \gamma'}$ and $G = G(\theta_0|\gamma_0)$ and $D = D(\theta_0|\gamma_0)$. Then, isolating $S$ in $G'W[GS +$

---

[1] The $vec(\bullet)$ operator stacks all columns of a matrix into a column vector, $vec(G) = (G_{1,1}, \ldots, G_{J,1}, G_{1,2}, \ldots, G_{J,2}, \ldots, G_{1,K}, \ldots, G_{J,K})'$.

$D] = 0$ gives

$$S = -(G'WG)^{-1}G'WD. \qquad (3)$$

Finally, because $g_n(\theta|\gamma)$, $G_n(\theta|\gamma)$ and $D_n(\theta|\gamma)$ converge uniformly in probability to their population counterparts together with $\text{plim}_{n\to\infty} \hat{\gamma} = \gamma_0$ and $\text{plim}_{n\to\infty} \hat{\theta} = \theta_0$, we have that $-(G'_n W_n G_n)^{-1} G'_n W_n D_n$ converges in probability to $-(G'WG)^{-1}G'WD$. This implies that the approximation in Corollary 1 in the main text converges in probability to the derivative $S = \frac{\partial \theta}{\partial \gamma'}$.

## C  Relation to Some Existing Sensitivity Measures

I here discuss some of the existing measures closest related to what I propose. The sensitivity measure proposed by Andrews, Gentzkow and Shapiro (2017) is related to the current measure. In particular, they propose to report $\hat{\Lambda}$ as a local measure of the sensitivity of $\theta$ to the included estimation moments in $g_n(\bullet)$. They do not consider the topic of the current paper and thus do not discuss sensitivity to calibrated parameters. The measure that I propose addresses this by weighting $\hat{\Lambda}$ by the effect of the calibrated parameters on each included moment through $D_n$ in equation Definition 1.

Another important contribution to the improvement of transparency is the recent work by Iskrev (2019). One of the sensitivity measures proposed in that study also measures how the estimated parameters are influenced by calibrated parameters. However, Iskrev (2019) focuses on Bayesian approaches and uses that the posterior distribution of $\theta$ and $\gamma$ is asymptotically jointly Normal to construct a local measure of sensitivity. Denote $\Sigma_\theta$ and $\Sigma_\gamma$ as the covariance matrices in the marginal asymptotic Normal distributions of $\theta$ and $\gamma$, respectively, and $\Sigma_{\theta,\gamma}$ as the covariance matrix between the two sets of parameters. From the asymptotic approximate Normal distribution, we have that the conditional mean vector of the estimated parameters, given the calibrated parameters, is

$$\mathbb{E}[\hat{\theta}|\hat{\gamma}] \stackrel{a}{=} \theta_0 + \Sigma_{\theta,\gamma} \Sigma_\gamma^{-1} (\hat{\gamma} - \gamma_0)$$

and Iskrev (2019) proposes the sensitivity measure

$$\Sigma_{\theta,\gamma} \Sigma_\gamma^{-1}. \qquad (4)$$

A drawback of this measure is, however, that it requires the calculation of the covariance matrix between $\theta$ and $\gamma$, $\Sigma_{\theta,\gamma}$. This covariance is, unfortunately, often not readily available in many empirical applications considered in the current study. If e.g. multiple data-sources or externally calibrated parameters are included in $\gamma$, calculating the covariance between $\gamma$ and $\theta$ is not straight forward – if not practically impossible.[2] Besides some of the references given in Footnote 1 in the main text, an example of such a situation is Gourinchas and Parker (2002). In that study, the Panel Study of Income Dynamics (PSID) is used to calibrate the exogenous income process in a first step and then subsequently estimate preference parameters using the Consumer Expenditure Survey (CEX), given the income process parameters. When the authors subsequently calculate standard errors of the estimated preference parameters, the authors assume that $\gamma$ and $\theta$ are uncorrelated, implying that $\Sigma_{\theta,\gamma} = 0$.

The measure in equation (4) is furthermore derived under the assumption that both sets of parameters are identified simultaneously, while mine is not. This is also evident from the linear regression example in the main text where $\beta_1$ and $\beta_2$ cannot be identified simultaneously from the one moment condition used. I view this as a strength of my approach because one motivation for using an externally calibrated $\gamma$ could be due to the unavailability of data that could identify $\gamma$.

There is also a literature focusing on *global* sensitivity measures of quantities of interest to model inputs. One approach could be to simulate values of $\gamma$ from some assumed distribution and investigate the resulting distribution of $\hat{\theta}$ from re-estimation of the model for each value of the drawn $\gamma$s.[3] These methods would often require re-estimation of the parameters relatively many times making such approaches computationally prohibitively expensive to apply to rich dynamic economic models, as I focus on here.

Recently, Harenberg, Marelli, Sudret and Winschel (2019) have proposed a *polynomial chaos expansion* to alleviate the computational burden associated with global sensitivity (or uncertainty quantification) approaches. However, building on series expansions, that approach also requires the re-estimation of the dynamic economic model at $M$ evaluation nodes to construct a global approximation of $\hat{\theta}$. If the dimension of the parameter space is large and/or the model complex, this can be quite computationally time demanding if a reasonable approximation is desired. Combining the local low-cost measure, that I propose, with the approach proposed in Harenberg, Marelli, Sudret and Winschel (2019) could potentially reduce the computational time required to perform global sensitivity analysis significantly: The local measure can guide researchers in which parameters are likely to require more evaluation nodes for a given degree of approximation accuracy.

---

[2] One strategy to uncovering *upper bounds* on the measure in (4) could be to use the worst-case upper bound on the co-variance structure following the approach suggested in Cocci and Plagborg-Møller (2019).

[3] See, e.g. Borgonovo and Plischke (2016) for a recent literature review.

# D  Model Implementation: Gourinchas and Parker (2002)

## D.1  Additional Figures and Tables

Table S1: Calibrated Parameters.

| $\sigma_n$ | $\sigma_u$ | $p$ | $\tilde{\omega}_{26}$ | $\sigma_{\omega_{26}}$ | $r$ |
|---|---|---|---|---|---|
| 0.0212 | 0.044 | 0.00302 | 0.061 | 1.784 | 0.0344 |

Figure S1: Income Growth and Family Shifter Calibration.

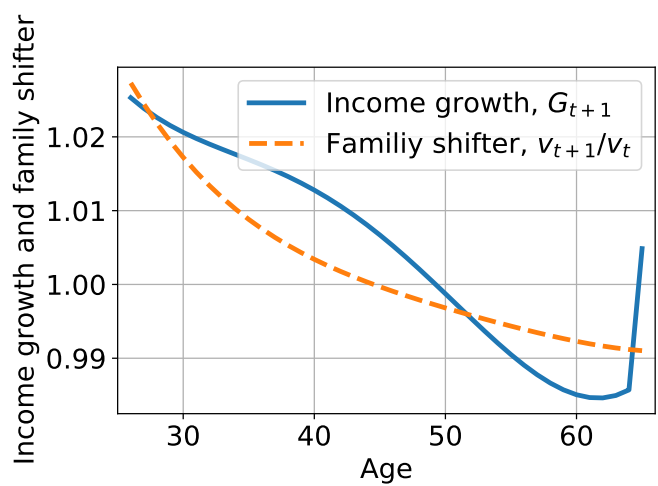

*Notes:* The figure shows the calibrated income growth, $G_{t+1}$, and the relative family shifter, $v(Z_{t+1})/v(Z_t)$.

Figure S2: Model Fit.

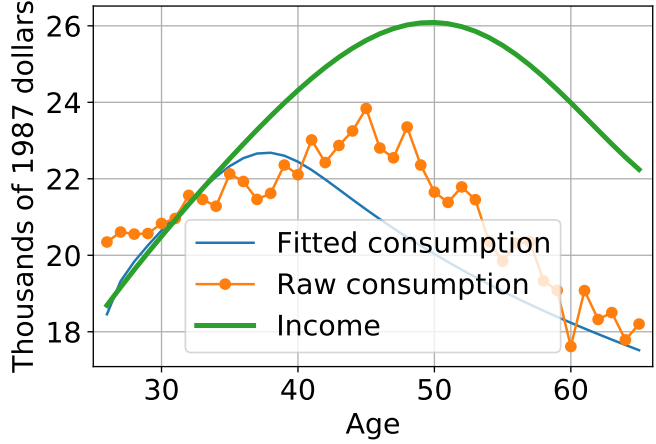

*Notes:* The figure illustrates the observed average income and consumption age profiles together with simulated average consumption from the re-estimated model.

Figure S3: Sensitivity of Parameter Estimates: Comparison of Derivatives.

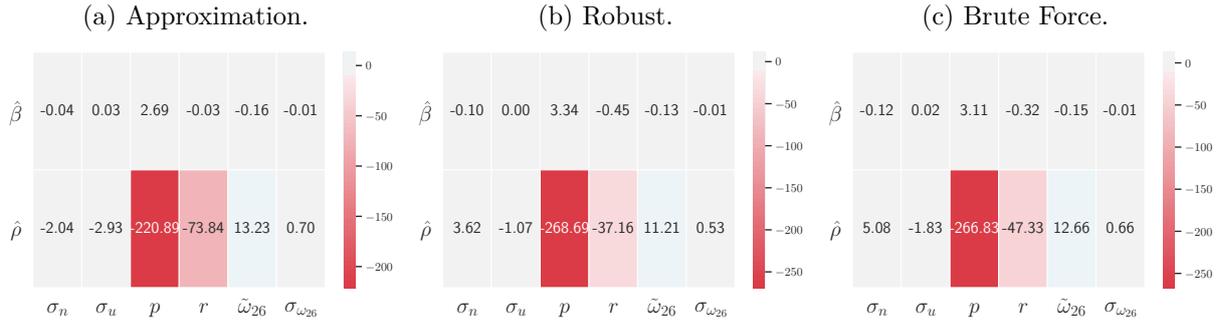

*Notes:* The figure illustrates the sensitivity of $\hat{\theta}$ with respect to the fixed parameters $\gamma$. The left panel shows the approximation of the derivative, the middle panel shows the general/robust derivative and the right panel shows the brute-force derivative calculated through re-estimation of the model.

## D.2 Solution Approach

As Gourinchas and Parker (2002), I use the Euler equation of a normalized model. From their Gauss code, and Appendix p.86 of the original paper, it seems that they assume that income is affected by the same taste shifter as consumption/utility. In the solution, I thus assume that $Y_{t+1} = v_{t+1}^{1/\rho} P_{t+1} U_{t+1}$. We can then normalize by $P_t v_t^{1/\rho}$ to get the Euler equation in normalized terms as

$$c_t^{-\rho} = \max\{m_t^{-\rho}, \beta(1+r)\mathbb{E}[(G_{t+1}N_{t+1}f_{t+1})^{-\rho} c_{t+1}^{-\rho}]\}$$

where $f_{t+1} = \left(\frac{v_{t+1}}{v_t}\right)^{\frac{1}{\rho}}$ adjusts for family composition. Normalized resources evolves according to

$$m_{t+1} = (1+r)a_t(G_{t+1}N_{t+1}f_{t+1})^{-1} + U_{t+1}$$

where $a_t = m_t - c_t$ is end-of-period normalized wealth. Constructing a grid of end-of-period wealth as $\vec{a}$ and approximating the expectation with two-dimensional Gauss-Hermite quadrature, optimal consumption can be found in closed form using the endogenous grid method (EGM) proposed by Carroll (2006) by inverting the Euler equation

$$c_t^\star = \left[(1+r)\beta \sum_{k=1}^{Q}\sum_{j=1}^{Q}[(G_{t+1}N^{(k)}f_{t+1})^{-\rho}\left(\check{c}_{t+1}^{(k,j)}\right)^{-\rho}]\right]^{-\frac{1}{\rho}}$$

where $\check{c}_{t+1}^{(k,j)} = \check{c}_{t+1}((1+r)(G_{t+1}N^{(k)}f_{t+1})^{-1}\vec{a} + U^{(j)})$ is the linearly interpolated next-period consumption for a given set of quadrature nodes $(k,j)$. The endogenous grid over resources is then $\vec{m}_t = \vec{a} + c_t^\star(\vec{m}_t)$. The credit constraint can be handled by including a lower point of $(m_{t+1}, c_{t+1}) = (0,0)$ when interpolating the next-period solution. I use 300 points in the $\vec{a}$ grid and $Q = 5$ quadrature nodes in each dimension. The implied consumption function is illustrated in Figure 1 with the calibrated parameters given below. The retirement consumption function is given by $c_{T+1}^\star(m_{T+1}) = \gamma_0 + \gamma_1 m_{T+1}$.

### D.3 Simulating Data

To simulate synthetic data (normalized by $v_t^{1/\rho}$), I draw $N_{sim} \times T$ standard normal shocks $\{\tilde{n}_{j,t}, \tilde{u}_{j,t}\}_{1,1}^{N_{sim},T}$ together with uniform draws $\{e_{j,t}\}_{1,1}^{N_{sim},T}$. I can then construct permanent and transitory income shocks, respectively, as

$$n_{j,t} = \exp(\sigma_n \tilde{n}_{j,t})$$
$$u_{j,t} = \exp(\sigma_u \tilde{u}_{j,t})(1-p)^{-1}\mathbf{1}(e_{j,t} > p)$$

I also draw standard normal initial wealth $\{\tilde{w}_{j,26}\}_1^{N_{sim}}$ and construct initial normalized resources as $m_{j,26} = \exp(\omega_{26} + \sigma_{\omega_{26}}\tilde{w}_{j,26}) + u_{j,26}$. Income is simulated as

$$P_{j,t} = \begin{cases} P_{26} & \text{if } t = 26 \\ G_t P_{j,t-1} n_{j,t} & \text{else} \end{cases}$$
$$Y_{j,t} = P_{j,t} u_{j,t}$$

and resources are

$$m_{j,t} = (1+r)(m_{j,t-1} - c_{j,t-1})(G_t n_{j,t} f_t)^{-1} + u_{j,t}$$

where consumption is found as the linearly interpolated optimal consumption solved above, $c_{j,t} = \check{c}_t(m_{j,t})$ and non-normalized consumption is then $C_{j,t} = c_{j,t} \cdot P_{j,t}$. All simulations are based on $N_{sim} = 500,000$ simulated individuals.

## D.4 Savings Motives Decomposition

Denote savings as the change in end-of-period wealth $s_{j,t} = A_{j,t} - A_{j,t-1}$. Solving and simulating an alternative model without income uncertainty and a modified retirement consumption rule, the life-cycle saving is defined as

$$s_{j,t}^{LC} = A_{j,t}^{LC} - A_{j,t-1}^{LC}$$

where the parameters of this model is $\sigma_n = \sigma_u = p = 0$ and $\gamma_1 = 0.0615$.[4] I also allow for borrowing in this version of the model up to 5 times the level of permanent income. The buffer saving is then given as $s_{j,t}^B = s_{j,t} - s_{j,t}^{LC}$. Figure S4 shows the average age profiles of these measures in the left panel and the average age profile of wealth split by life cycle and buffer wealth in the right panel.

Figure S4: Savings Motives Decomposition.

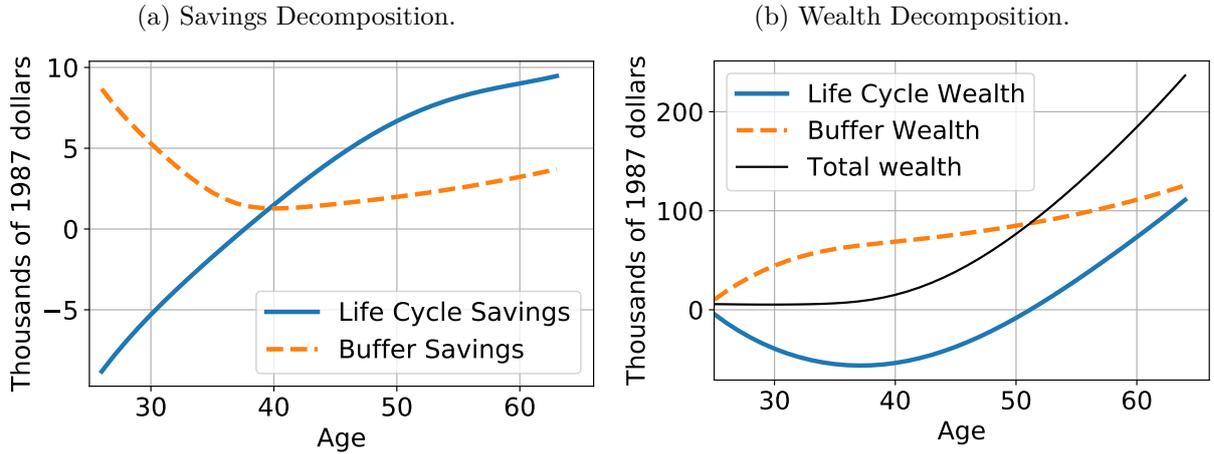

*Notes:* The left panel illustrates the saving decomposition into life cycle savings and buffer savings, comparable to the top panel of figure 7 in Gourinchas and Parker (2002). The right panel illustrates the wealth decomposition into life cycle wealth and buffer wealth, comparable to the bottom panel of figure 7 in Gourinchas and Parker (2002).

# E  Additional Figures and Tables Oswald (2019)

---

[4] The retirement consumption function is modified such that there is full certainty after retirement,

$$\gamma_1 = \frac{1 - \hat{\beta}^{\frac{1}{\rho}}(1+r)^{\frac{1}{\rho}-1}}{1 - \left(\hat{\beta}^{\frac{1}{\rho}}(1+r)^{\frac{1}{\rho}-1}\right)^{D-T}}$$

where $D = 88$ and $T = 65$.

Table S2: Parameters in Oswald (2019).

| Estimated parameters in $\theta$ | | value | Calibrated parameters in $\gamma$ | | value |
|---|---|---|---|---|---|
| *Utility function* | | | CRRA coefficient | $\tilde{\gamma}$ | 1.43 |
| Owner premium size 1 | $\xi_1$ | $-0.009$ | Discount factor | $\beta$ | 0.96 |
| Owner premium size 2 | $\xi_2$ | 0.003 | AR(1) of pers. inc. shock | $\rho$ | 0.96 |
| Util. of cons. scale | $\eta$ | 0.217 | Std. of pers. inc. shock | $\sigma$ | 0.118 |
| Continuation value | $\omega$ | 4.364 | Transaction cost | $\phi$ | 0.06 |
| *Moving costs* | | | Down-payment proportion | $\chi$ | 0.20 |
| Constant | $\alpha_0$ | 3.165 | Risk-free interest rate | $r$ | 0.04 |
| Age | $\alpha_1$ | 0.017 | 30-year mortgage rate | $r^m$ | 0.055 |
| Age$^2$ | $\alpha_2$ | 0.0013 | | | |
| Owner | $\alpha_3$ | 0.217 | | | |
| Household size | $\alpha_4$ | 0.147 | | | |
| Proportion of high type | $\pi_\tau$ | 0.697 | | | |
| *Amenities* | | | | | |
| New England | $A_{NwE}$ | 0.044 | | | |
| Middle Atlantic | $A_{MdA}$ | 0.112 | | | |
| Middle Atlantic | $A_{StA}$ | 0.168 | | | |
| West North Central | $A_{WNC}$ | 0.090 | | | |
| West South Central | $A_{WSC}$ | 0.122 | | | |
| East North Central | $A_{ENC}$ | 0.137 | | | |
| East South Central | $A_{ESC}$ | 0.063 | | | |
| Pacific | $A_{pcf}$ | 0.198 | | | |
| Mountain | $A_{Mnt}$ | 0.124 | | | |

Table S3: Sensitivity of Parameters in Oswald (2019). Elasticities.

|  | $\tilde{\gamma}$ | $\beta$ | $\rho$ | $\sigma$ | $\phi$ | $\chi$ | $r$ | $r^m$ |
|---|---|---|---|---|---|---|---|---|
| $\hat{\xi}_1$ | 114.817 | -86.300 | -189.804 | -3.443 | 1.378 | -1.868 | -0.179 | -0.190 |
| $\hat{\xi}_2$ | -1050.357 | 209.728 | 2772.968 | -2.366 | -5.134 | 37.807 | 1.255 | 1.537 |
| $\hat{\eta}$ | -36.763 | 10.532 | 232.386 | -2.831 | -5.039 | -1.397 | -0.023 | 0.069 |
| $\hat{\omega}$ | 0.165 | 0.162 | -0.047 | -0.003 | -0.023 | -0.006 | -0.001 | 0.000 |
| $\hat{\alpha}_0$ | -1.174 | -0.019 | 2.211 | 0.010 | 0.013 | 0.023 | 0.001 | -0.000 |
| $\hat{\alpha}_1$ | -85.165 | -32.862 | 547.621 | -1.296 | -7.673 | 0.002 | -0.122 | -0.028 |
| $\hat{\alpha}_2$ | 1554.699 | 201.729 | -3620.681 | -21.701 | 18.066 | -10.432 | -0.093 | -2.910 |
| $\hat{\alpha}_3$ | -0.936 | 3.430 | 20.203 | 0.129 | 0.015 | 0.016 | 0.008 | 0.005 |
| $\hat{\alpha}_4$ | 37.091 | -3.228 | -17.754 | -0.277 | -0.088 | -0.977 | -0.058 | 0.033 |
| $\hat{\pi}_\tau$ | 0.057 | -0.003 | -0.057 | -0.000 | -0.003 | -0.002 | -0.000 | 0.000 |
| $\hat{A}_{NwE}$ | 62.353 | -0.490 | -75.947 | 0.373 | 2.561 | -1.728 | 0.008 | -0.134 |
| $\hat{A}_{MdA}$ | -127.703 | 11.875 | 437.474 | -1.113 | -5.718 | 2.360 | 0.088 | 0.019 |
| $\hat{A}_{StA}$ | -29.806 | 3.227 | -9.579 | 0.042 | -0.107 | 0.731 | 0.024 | -0.032 |
| $\hat{A}_{WNC}$ | -237.423 | 6.282 | -21.925 | 3.446 | 2.524 | 8.880 | 0.379 | -2.884 |
| $\hat{A}_{WSC}$ | 56.868 | 0.880 | -54.668 | 0.416 | 0.827 | -0.892 | -0.024 | 0.050 |
| $\hat{A}_{ENC}$ | -4.929 | 0.411 | 12.863 | -0.049 | -0.035 | 0.074 | -0.004 | 0.002 |
| $\hat{A}_{ESC}$ | 214.474 | 18.102 | -1198.287 | -2.251 | 12.797 | -4.375 | 0.642 | 0.463 |
| $\hat{A}_{Pcf}$ | -2.024 | 0.576 | 20.273 | 0.215 | 0.228 | 0.203 | -0.008 | 0.019 |
| $\hat{A}_{Mnt}$ | 281.108 | 131.541 | -2120.151 | 37.155 | -1.200 | 12.639 | 0.665 | -0.504 |

*Notes*: The table reports the sensitivity of the estimated $\theta$ parameters in Oswald (2019). Elasticities are reported.